\documentclass[12pt,preprint]{aastex62}

\usepackage{graphicx}

\newcommand{\lya}{Ly$\alpha$}
\newcommand{\Ha}{H$\alpha$}

\newcommand{\FeII}{Fe~{\sc ii}}
\newcommand{\FeIII}{Fe~{\sc iii}}

\newcommand{\MgII}{Mg~{\sc ii}}

\newcommand{\OIII}{[O~{\sc iii}]}
\newcommand{\kms}{\hbox{km~s$^{-1}$}}
\newcommand{\cmsq}{\hbox{cm$^{-2}$}}

\newcommand{\flux}{\hbox{erg~cm$^{-2}$~s$^{-1}$}}

\newcommand{\pflux}{\hbox{photon~cm$^{-2}$~s$^{-1}$}}

\newcommand{\lumin}{\hbox{erg~s$^{-1}$}}

\newcommand{\nh}{\hbox{${N}_{\rm H}$}}

\newcommand{\be}{\begin{equation}}
\newcommand{\ee}{\end{equation}}
\newcommand{\ba}{\begin{eqnarray}}
\newcommand{\ea}{\end{eqnarray}}

\newcommand{\rosat}{\emph{ROSAT}}

\newcommand{\swift}{\emph{Swift}}

\newcommand{\fermi}{\emph{Fermi}}
\newcommand{\simgt}{\lower 2pt \hbox{$\, \buildrel {\scriptstyle >}\over {\scriptstyle\sim}\,$}}
\newcommand{\simlt}{\lower 2pt \hbox{$\, \buildrel {\scriptstyle <}\over {\scriptstyle\sim}\,$}}

\newcommand{\ls}{\lower 2pt \hbox{$\;\scriptscriptstyle \buildrel<\over\sim\;$}}
\newcommand{\gs}{\lower 2pt \hbox{$\;\scriptscriptstyle \buildrel>\over\sim\;$}}

\newcommand{\btwo}{B2\,1420+32}

\graphicspath{{./}{figures/}}

\received{}
\revised{}
\accepted{}

\shorttitle{The Changing Look Blazar B2 1420+32}
\shortauthors{Mishra et al.}

\begin{document}
\title{The Changing Look Blazar B2 1420+32}

\correspondingauthor{Hora D. Mishra}
\email{hora.mishra@ou.edu}
\author[0000-0002-6821-5927]{Hora D. Mishra}
\affil{Homer L.\ Dodge Department of Physics and Astronomy,
University of Oklahoma, Norman, OK 73019, USA}

\author[0000-0001-9203-2808]{Xinyu Dai}
\affil{Homer L.\ Dodge Department of Physics and Astronomy,
University of Oklahoma, Norman, OK 73019, USA}
\email{xdai@ou.edu}

\author[0000-0003-0853-6427]{Ping Chen}
\affiliation{Kavli Institute for Astronomy and Astrophysics, Peking University, Yi He Yuan Road 5, Hai Dian District, Beijing 100871, China}
\affiliation{Department of Astronomy, School of Physics, Peking University, Yi He Yuan Road 5, Hai Dian District, Beijing 100871, China}

\author{Jigui Cheng}
\affil{School of Physical Science and Technology, Guangxi University, Nanning 530004, China}

\author{T. Jayasinghe}
\affiliation{Department of Astronomy, The Ohio State University, 140 West 18th Avenue, Columbus, OH 43210, USA}

\author[0000-0002-2471-8442]{Michael A. Tucker}
\altaffiliation{DOE CSGF Fellow}
\affiliation{Institute for Astronomy, University of Hawai\`{}i at Manoa, 2680 Woodlawn Dr., Honolulu, HI 96822, USA}

\author[0000-0001-5661-7155]{Patrick J. Vallely}
\altaffiliation{NSF Graduate Research Fellow}
\affiliation{Department of Astronomy, The Ohio State University, 140 West 18th Avenue, Columbus, OH 43210, USA}

\author[0000-0001-7485-3020]{David Bersier}
\affiliation{Astrophysics Research Institute, Liverpool John Moores University,  146 Brownlow Hill, Liverpool L3 5RF, UK}

\author{Subhash Bose}
\affiliation{Department of Astronomy, The Ohio State University, 140 West 18th Avenue, Columbus, OH 43210, USA}
\affiliation{Center for Cosmology and AstroParticle Physics, The Ohio State University, 191 W.\ Woodruff Ave., Columbus, OH 43210, USA}

\author{Aaron Do}
\affiliation{Institute for Astronomy, University of Hawai\`{}i at Manoa, 2680 Woodlawn Dr., Honolulu, HI 96822, USA}

\author{Subo Dong}
\affiliation{Kavli Institute for Astronomy and Astrophysics, Peking University, Yi He Yuan Road 5, Hai Dian District, Beijing 100871, China}

\author[0000-0001-9206-3460]{Thomas~W.-S.~Holoien}
\affiliation{The Observatories of the Carnegie Institution for Science, 813 Santa Barbara St., Pasadena, CA 91101, USA}

\author[0000-0003-1059-9603]{Mark E. Huber}
\affiliation{Institute for Astronomy, University of Hawai\`{}i at Manoa, 2680 Woodlawn Dr., Honolulu, HI 96822, USA}

\author[0000-0001-6017-2961]{Christopher~S.~Kochanek}
\affiliation{Department of Astronomy, The Ohio State University, 140 West 18th Avenue, Columbus, OH 43210, USA}
\affiliation{Center for Cosmology and AstroParticle Physics, The Ohio State University, 191 W.\ Woodruff Ave., Columbus, OH 43210, USA}

\author{Enwei Liang}
\affil{School of Physical Science and Technology, Guangxi University, Nanning 530004, China}

\author[0000-0003-3490-3243]{Anna V. Payne}
\altaffiliation{NASA Fellow}
\affiliation{Institute for Astronomy, University of Hawai\`{}i at Manoa, 2680 Woodlawn Dr., Honolulu, HI 96822, USA}

\author{Jose Prieto}
\affiliation{N\'ucleo de Astronom\'ia de la Facultad de Ingenier\'ia y Ciencias, Universidad Diego Portales, Av. Ej\'ercito 441, Santiago, Chile}

\author[0000-0003-4631-1149]{Benjamin J. Shappee}
\affiliation{Institute for Astronomy, University of Hawai\`{}i at Manoa, 2680 Woodlawn Dr., Honolulu, HI 96822, USA}

\author{K.~Z.~Stanek}
\affiliation{Department of Astronomy, The Ohio State University, 140 West 18th Avenue, Columbus, OH 43210, USA}
\affiliation{Center for Cosmology and AstroParticle Physics, The Ohio State University, 191 W.\ Woodruff Ave., Columbus, OH 43210, USA}

\author[0000-0002-9044-9383]{Saloni Bhatiani}
\affil{Homer L.\ Dodge Department of Physics and Astronomy,
University of Oklahoma, Norman, OK 73019, USA}

\author{John Cox}
\affil{Homer L.\ Dodge Department of Physics and Astronomy,
University of Oklahoma, Norman, OK 73019, USA}

\author[0000-0002-1650-7936]{Cora DeFrancesco}
\affil{Homer L.\ Dodge Department of Physics and Astronomy,
University of Oklahoma, Norman, OK 73019, USA}

\author{Zhiqiang Shen}
\affil{Shanghai Astronomical Observatory, Chinese Academy of Sciences, 80 Nandan Road, Shanghai 200030, China}

\author[0000-0003-2377-9574]{Todd~A.~Thompson}
\affiliation{Department of Astronomy, The Ohio State University, 140 West 18th Avenue, Columbus, OH 43210, USA}
\affiliation{Center for Cosmology and AstroParticle Physics, The Ohio State University, 191 W.\ Woodruff Ave., Columbus, OH 43210, USA}

\author[0000-0003-4874-0369]{Junfeng Wang}
\affil{Department of Astronomy, Xiamen University, Xiamen, Fujian 361005, China}

\begin{abstract}
Blazars are active galactic nuclei with their relativistic jets pointing toward the observer, with two major sub-classes, the flat spectrum radio quasars and BL Lac objects.   
We present multi-wavelength photometric and spectroscopic monitoring observations of the blazar, B2 1420+32,
focusing on its outbursts in 2018-2020. Multi-epoch spectra show that the blazar exhibited large scale spectral variability in both its continuum and line emission, accompanied by dramatic gamma-ray and optical variability by factors of up to 40 and 15, respectively, on week to month timescales.  
Over the last decade, the gamma-ray and optical fluxes increased by factors of 1500 and 100, respectively.
\btwo\ was an FSRQ with broad emission lines in 1995. Following a series of flares starting in 2018, it transitioned between BL Lac and FSRQ states multiple times, with the emergence of a strong Fe pseudo continuum. 
Two spectra also contain components that can be modeled as single-temperature black bodies of 12,000 and 5,200~K.
Such a collection of ``changing look'' features has never been observed previously in a blazar.
We measure $\gamma$-ray-optical and the inter-band optical lags implying emission region separations of less than 800 and 130 gravitational radii respectively.
Since most emission line flux variations, except the Fe continuum, are within a factor of 2--3, the transitions between FSRQ and BL Lac classifications are mainly caused by the continuum variability.
The large Fe continuum flux increase suggests the occurrence of dust sublimation releasing more Fe ions in the central engine and an energy transfer from the relativistic jet to sub-relativistic emission components. 
\end{abstract}

\keywords{black hole physics --- (galaxies:) quasars: emission lines --- (galaxies:) quasars: general
(galaxies:) quasars: individual (B2\,1420+32)}

\section{Introduction} \label{sec:intro}
Active Galactic Nuclei (AGN) are subdivided into several broad categories. Type I AGNs (also called quasars, Seyfert I) show a blue continuum from an accretion disk and broad emission lines created by photoionization. The continuum flux stochastically varies with modest amplitudes \citep[e.g.,][]{macleod10} and the broad lines respond after a delay. Type II AGNs (or Seyfert II) show only narrow lines and no continuum variability \citep[e.g.,][]{khachikian1974, nagao2001,peterson04}.  The most common paradigm to unify the two classes is to assume that the line of sight to the central engine is unobscured for Type I AGN and obscured for Type II AGN \citep[e.g.,][]{antonucci1993,up95}. Most AGNs are not strong radio sources (i.e., ``radio quiet''). Those which are radio loud can be divided into flat and steep spectrum radio sources. Here, the radio emission is believed to be due to a jet. The emission from flat spectrum radio quasars (FSRQs) is dominated by direct emission from the jet \citep[e.g.,][]{garofalo2018} and the steep spectrum sources are dominated by emission from the extended ``lobes'', where the jet is interacting with the ambient medium \citep[e.g.,][]{fanti1990}. Since the jets are relativistic, emission from the jet can dominate if the jet is pointed towards the observer. In the extreme case of blazars (also optically violent variables, OVVs), the jet emission dominates at all wavelengths and no emission lines from an underlying quasar are visible. Blazars also show much higher amplitude and shorter time scale variability than quasars at all wavelengths, from the radio band to $\gamma$-rays \citep[e.g.,][]{edelson1987, up95, sesar2007}.

An increasingly powerful means of understanding these divisions is to discover and analyze ``changing look'' AGN, where a source moves from one class to another. Most examples are AGN shifting between Type I and Type II spectra \citep[e.g.,][]{matt2003, bianchi2005, marchese2012, shappee14}, a change which calls into question the standard unification picture for the difference between these classes. With the availability of large spectroscopic and time domain surveys, there have been a series of systematic searches that have found increasing number of examples of such AGNs \citep[e.g.,][]{ac16, kollatschny2018, ai2020}. One interesting bias of these searches is that they generally exclude blazars from the search because their optical variability amplitudes are so high. 
This is unfortunate, because changing look phenomena in blazars can provide useful insight into understanding the origin and particle acceleration processes of the radio jets, the role of changing structure and geometry of the jets, and the accretion disk-jet connection \citep[e.g.,][]{falcke1995}. Jets are also an important feedback mechanism at the galaxy cluster scale \citep[e.g.,][]{mn12} and the galaxy scale for the milder decelerated jets in radio galaxies \citep[][]{capetti05,ishibashi14,baldi19}.  

Blazars can be broadly divided into two categories -- FSRQs and BL Lac objects, based on the rest-frame equivalent width of the strongest broad emission line. If the equivalent width is less than 5\AA, the blazar is classified as a BL Lac object, otherwise, an FSRQ \citep{up95}.
An alternate classification is based on the total broad line luminosity in units of the Eddington luminosity with the boundary at $\sim 10^{-3} L_{Edd}$ \citep{ghisellini11}. As with Type I and II AGNs, there are arguments about potential unification schemes for the two classes.
The broadband spectral energy distributions (SED) of blazars have two peaks. There is a lower energy component from sub-millimeter to X--ray energies due to synchrotron emission and a high energy component at MeV to TeV energies due to the inverse Compton process.
\citet{fossati98} and \citet{donato01} proposed that FSRQs and BL Lacs are a sequence, where the broadband SED moves blue-ward, with the bolometric luminosity decreasing from FSRQS to BL Lacs because cooling is more efficient in FSRQ jets than in BL Lacs \citep{ghisellini98}. In this picture, FSRQs have an efficient accretion disk powering the broad-line region (BLR) and BL Lacs have inefficient accretion disks.
Alternative unification schemes for FSRQs and BL Lacs have been proposed \citep[e.g.,][]{giommi12}.

Studying systems that alternate between FSRQ and BL Lac states, using photometric and spectroscopic data, should illuminate their differences, but there are few studies in literature that can really be used to explore the question of blazar unification. While there are many studies of blazar variability at particular energies as well as studies of correlations of the variability between different energies and changes in the overall SED \citep[][]{paliya2015, zhang2015, yoo2020}, there are many fewer spectroscopic monitoring studies of blazars \citep[e.g.,][]{zb86, bregman86, perez89, vermeulen95, ulrich97, corbett00}. In a few cases, large emission line flux changes have been observed. For example, \citet{vermeulen95} reported an increase in the \Ha\ luminosity by a factor of 10 for the BL Lacertae prototype VRO 42.22.01 between 1989 and 1995. More recently, \citet{isler13, isler15} observed BEL equivalent width changes accompanied by large Fermi $\gamma$-ray flares for four FSRQs.

Here we discuss observations of the blazar B2~1420$+$32 doing this not once, but multiple times over a two year period.
B2~1420$+$32 at $z=0.682$ was identified as an FSRQ and has been detected from radio to $\gamma$--energies. From its luminosity and broad line width, \citet{brotherton15} estimated a black hole mass of $M_{BH} \simeq 4 \times 10^8 M_\odot$, corresponding to a minimum light crossing time of approximately $r_g/c=GM_{BH}/c^3 = 0.5$~hours (rest-frame). We first became interested in the source after the All-Sky Automated Survey for Supernovae \citep[ASAS-SN,][]{shappee14, kochanek17} detected an optical flare of $> 2$~mag on 2017 Dec 28 \citep{{stanek17}} after nearly a decade of relative quiescence in the Catalina Real Time Survey \citep[CRTS,][]{drake2009}. At this point we started to obtain multi-color light curves using the Las Cumbres Observatory Global Telescope Network \citep[LCOGT,][]{Brown+2013} 1m telescopes and spectra from a variety of sources. Over the next two years, additional flares were flagged in the optical/near-IR \citep[e.g.,][]{carrasco19,marchini19}, $\gamma$~rays \citep{ciprini18}, and even very high energy (VHE, $E>100$~GeV) $\gamma$-rays \citep{mirzoyan20}.

Here we report the results of our campaign. The most striking result is that during these high amplitude brightness fluctuations, B2~1420$+$32 shifted back and forth between the optical spectrum of an FSRQ and that of a BL Lac several times, while also developing new spectral features. We discuss the photometric data in section \ref{sec:pho}, including cross correlation analyses between the various energy bands. We present and discuss the spectral evolution in section \ref{sec:spec}. We consider the implications of this behavior for understanding FSRQs, blazars and their differences in section \ref{sec:dis}. We adopt the cosmological parameters $\Omega_{m} = 0.3$, $\Omega_{\Lambda} = 0.7$, and $H_0 = 70$~km~s$^{-1}$~Mpc$^{-1}$.

\begin{figure}
    \centering
    \centerline{\includegraphics[width=1.20\textwidth]{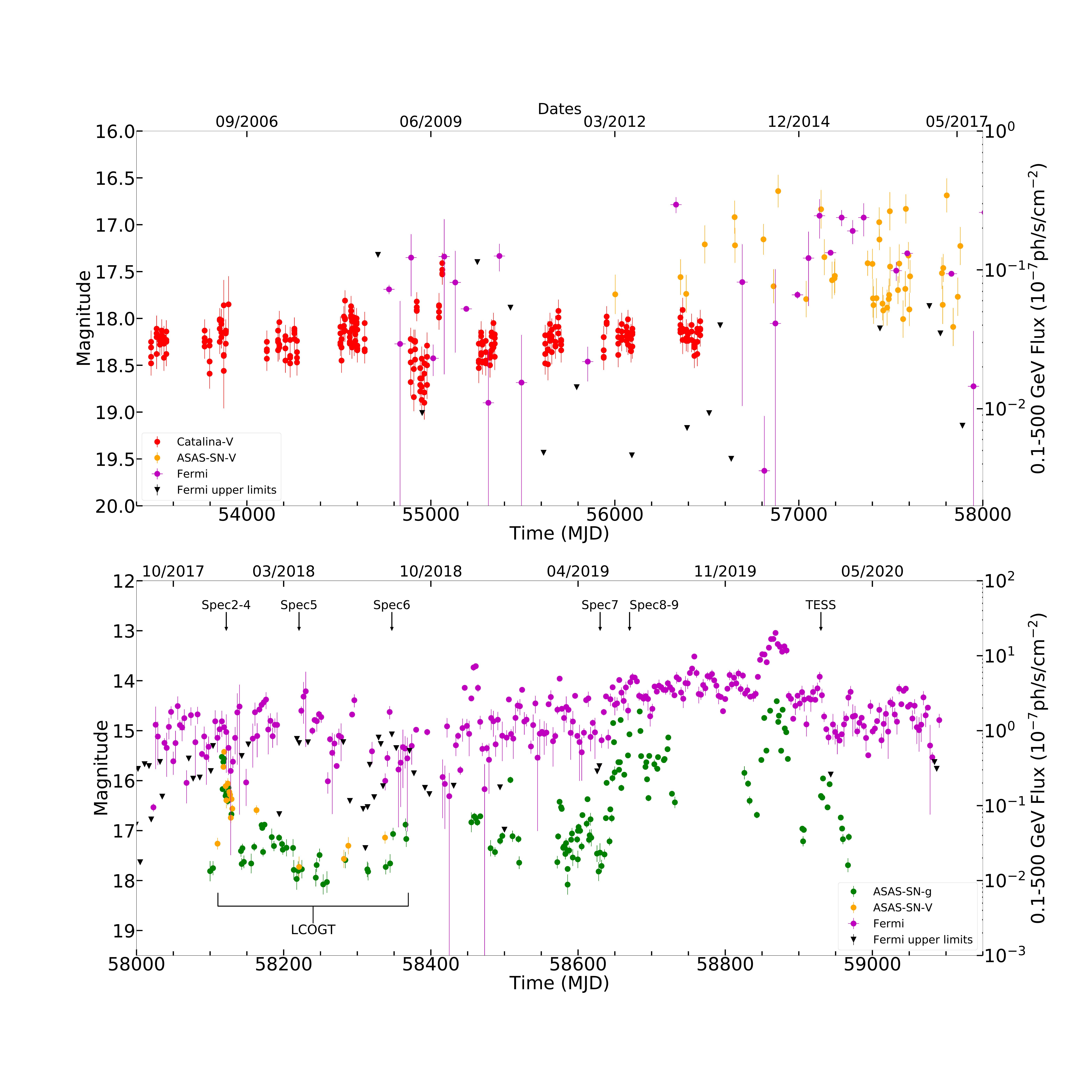}}
    \caption{Long-term optical and 0.1--500~Gev $\gamma$-ray light curves of \btwo, where the bottom panel shows the enhanced optical and $\gamma$-ray activities between MJD~58000 and 59100 and the top panel covers the range before MJD~58000.  The $\gamma$-ray light curve is binned by 3 days in the bottom panel and 2 months in the top panel. The epochs of the spectroscopic observations, LCOGT, and TESS observations are marked.}
    \label{fig:lc}
\end{figure}

\begin{figure}
    \centering
    \centerline{\includegraphics[width=1.0\textwidth]{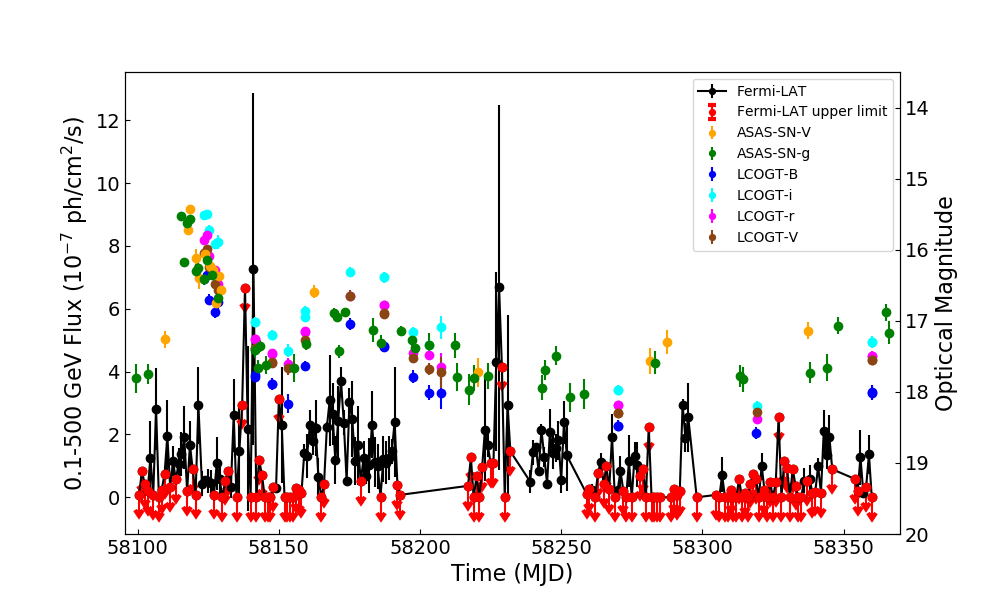}}
    \centerline{\includegraphics[width=1.0\textwidth]{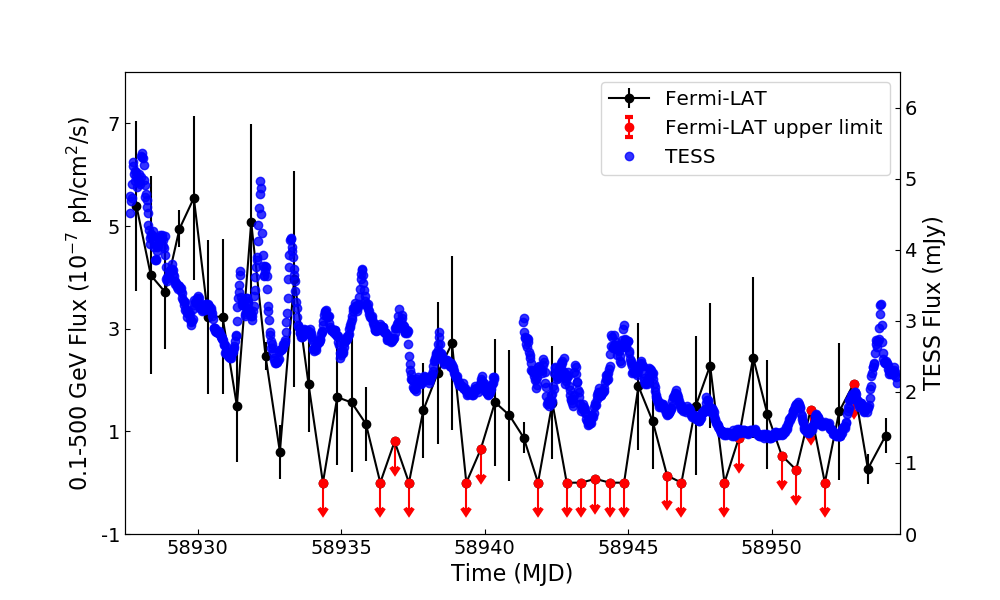}}
    \caption{Multi-band optical LCOGT, ASAS-SN, and Fermi light curves of \btwo\ between MJD 58124 and 58360, where the Fermi data is binned by 1 day (top). TESS and Fermi light curves of \btwo\ between MJD 58927 and 58954, where the TESS cadence is 30 min and we binned the Fermi data by 0.5 days (bottom). The optical flux shows many flares on sub-day timescales with amplitudes exceeding 50$\%$.}
    \label{fig:tess}
\end{figure}

\begin{figure}
    \centering
    \centerline{\includegraphics[width=0.6\textwidth]{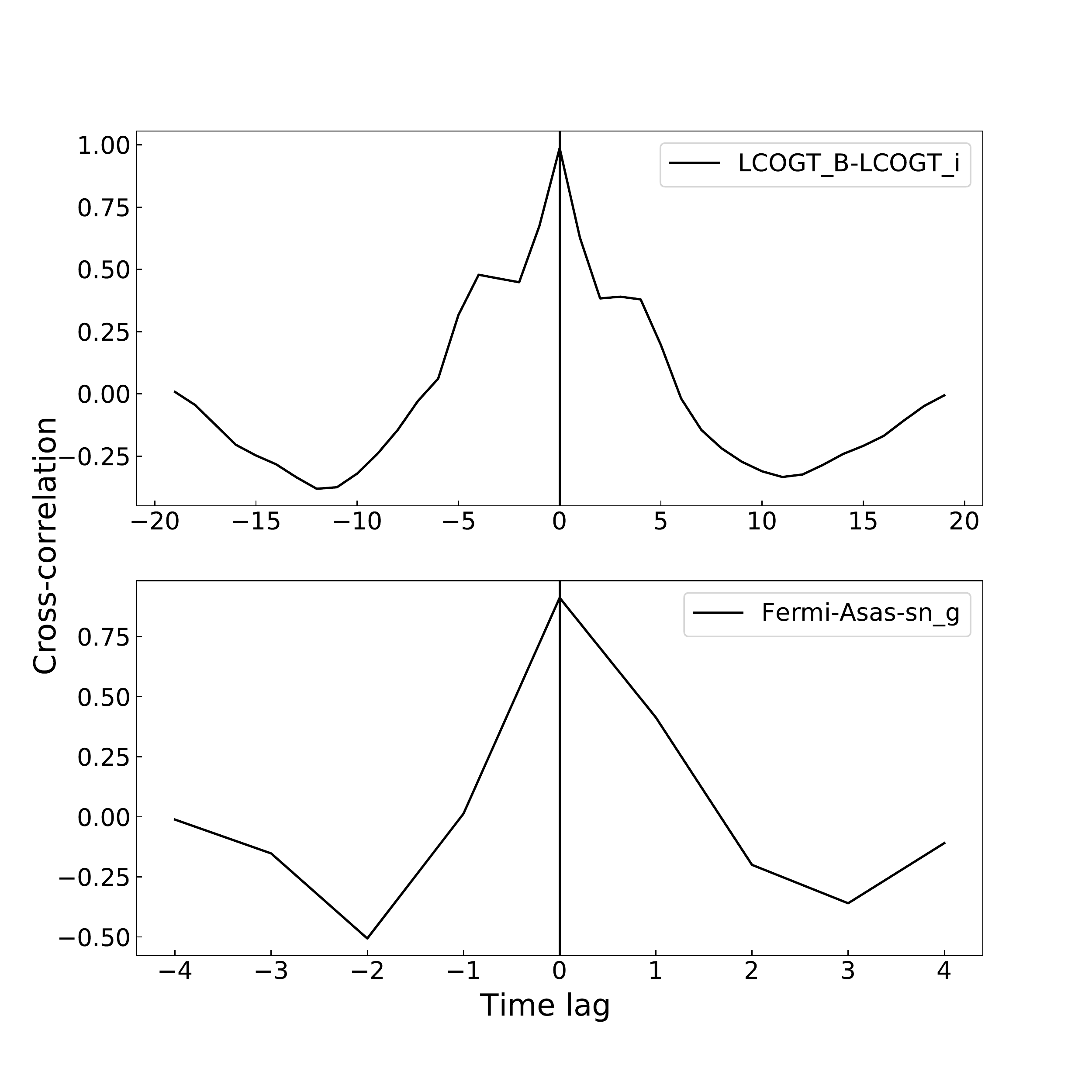}}
    \caption{Two examples of the cross-correlations between the LCOGT B and LCOGT i-band (top panel) and the Fermi and ASAS-SN g-band light curves-overall peak (bottom panel) as a function of the time lag. For these cross-correlations, the lag estimates are 0.0$\pm$0.4 days (top) and 0.0$\pm$0.5 days (bottom).}
    \label{fig:iccf}
\end{figure}

\section{Temporal Evolution \label{sec:pho}}
\btwo\ is detected across the entire electromagnetic spectrum -- from the radio to $\gamma$-ray bands. In this section we examine the optical and $\gamma$-ray variability of \btwo\ and the
correlations between them. We also obtained a single Swift \citep{gehrels04} XRT \citep{burrows2005} X-ray observation. The X-ray observation is only 
described in the text. 

For the $\gamma$-rays, we analyzed the full 12-year \fermi-LAT \citep{atwood09} PASS8 data in the 0.1--500~GeV band from MJD~54689 to 59090. We used different temporal bins depending on the brightness of the source. Prior to MJD~58000, we used bins of two months, and afterwards we used bins of 3 days. During the period
with LCOGT monitoring data, we used bins of a single day, and during the TESS observations we used bins of 0.5~days to better match the high cadence TESS optical data.
For each bin, we performed a maximum likelihood analysis using the \textit{PYTHON} script \textit{make4FGLxml.py}\footnote{\url{https://fermi.gsfc.nasa.gov/ssc/data/analysis/user/}} to model the source spectrum and flux. The minimum detection threshold is set at $TS=2.69$, corresponding to the $90\%$ confidence level.

The optical data came from multiple sources. The earliest data is a V-band light curve from CRTS \citep{drake2009}. Next we used the ASAS-SN V- and g-band data \citep{shappee14,kochanek17}, with the light curves obtained using image subtraction as described in \citet{Jayasinghe2018} and \citet{Jayasinghe2019b}. 
B2~1420$+$32 was observed by the Transiting Exoplanet Survey Satellite \citep[TESS,][]{ricker+2015} during Sector 23. The TESS band light curve was extracted using image subtraction methods optimized for TESS, as described in \cite{vallely2019}.

We monitored B2~1420$+$32 in the B, V, r and i-bands with the LCOGT \citep{Brown+2013} 1-m telescope at the McDonald Observatory. After basic reduction, the images were downloaded from Las Cumbres Observatory Science Archive (https://archive.lco.global). We used the IRAF \citep{Tody1986} apphot task to perform aperture photometry with an aperture size twice the full width at half maximum (FWHM) of the stellar profile. AAVSO Photometric All-Sky Survey \citep[APASS;][]{Henden+2016} DR9 catalog stars were used for photometric calibration. 

We observed \btwo\ with \swift\ \citep{gehrels04} on 2018-01-20:UT15:44:46 for 1~ks with the XRT in the WT mode.  We measured a net count rate of $0.046\pm 0.019$~ct~s$^{-1}$ in the 0.2--10~keV band with an unabsorbed flux of $1.7\times10^{-12}$~\flux, assuming a powerlaw photon index of 1.7 and adopting a Galactic absorption of \nh$=1.07\times10^{20}$~\cmsq\ \citep{hi4pi15}. Compared to the \rosat\ All-Sky Survey flux of $5.4\times10^{-13}$~\flux\ in the 0.1--2.4~keV band \citep{massaro09} or equivalently $9.3\times10^{-13}$~\flux\ in the 0.2--10~keV band, the X-ray flux increased by a factor of 2.
\citet{prince19} measured an increase in X-ray activity using the Swift-XRT on MJD 58830 with an unabsorbed flux of $8.3\pm1.7\times10^{-12}$ \flux\ in the 0.3--10~keV band, assuming a powerlaw photon index of $1.35\pm0.22$. This is a further increase in the XRT flux by a factor of 5.

Figures \ref{fig:lc} and \ref{fig:tess} show the optical and $\gamma$-ray evolution on a series of time scales. Fig.\ref{fig:lc} shows the evolution over the last 15 years and is divided roughly into the pre and post-outburst phases. Over the entire decade long period of photometric monitoring, the long-term $\gamma$-flux increased by a factor of 1500, when comparing the highest and lowest $\gamma$-ray fluxes detected, and the optical flux increased by a factor of 100. The top panel of Fig.\ref{fig:tess} shows the period in 2018 where we obtained the higher cadence, multi-band LCOGT data. Finally, the bottom panel of Fig.\ref{fig:tess} shows the 26 day period of TESS observations in early 2020.
For roughly the decade prior to the ASAS-SN flare at the end of 2017, the source was fairly quiescent. The optical mean magnitude was $\langle V \rangle = 18.3$ with a scatter of $0.20$~mag. This is similar to its fluxes in the SDSS survey, measured on March 17, 2004 and the early CRTS data. Other than a weak flare in June 2009 there is little variability.
Similarly, the $\gamma$-ray flux is low (mean = 1.4$\times10^{-8}$ ph/s/cm$^2$), with too few counts to really characterize the variability.

The bottom panel of Fig.\ref{fig:lc} shows that the optical flare flagged by ASAS-SN was accompanied by a $\gamma$-ray flare (MJD 58100--58150). The optical outburst, where the flux increases by 3.0$\pm$0.2 mags (factor of 16), was intensely followed-up with multi-band LCOGT observations (Figure~\ref{fig:tess}, top panel). However, the $\gamma$-ray flux does not show a significant flare at the peak of the optical outburst (MJD 58125). 
Afterwards, the optical flux remained somewhat higher than before the flare, but the $\gamma$-ray flux increases by 2.7$\pm$0.3 mags.
Another $\gamma$-ray flare is observed near November 2018 (MJD 58450) when the flux increases by a factor of 16, after which the $\gamma$-ray flux stays in the high state, and the amplitudes of the $\gamma$-ray flares are reduced to 1.5~mag.
Then, near May 2019, both the optical and $\gamma$-ray fluxes increased by another two orders of magnitude (Fig.\ref{fig:lc}, bottom panel), with a further increase in March 2020 (MJD~58868) to a peak of $g = 14.4$ mags and 2.0$\times10^{-6}$ ph/s/cm$^2$, respectively. At this peak, the optical flux is 6 times brighter than the pre-flare mean, while the $\gamma$-ray flux is 16 times brighter. The optical and $\gamma$-ray data follow each other almost exactly.

Finally, the bottom panel of Figure~\ref{fig:tess} shows the brief period of TESS observations (MJD 58927--58954) with the Fermi data binned into 0.5~day intervals. The high S/N and cadence (0.5 hours) TESS light curve shows multiple, intra-day flares with the flux change being as large as a factor of 4, for example, at MJD 58927 and 58933. The $\gamma$-flux appears to track the TESS light curve (Figure~\ref{fig:tess}, bottom panel), although the lower S/N in the smaller temporal bins limits the comparison. 
While the high amplitude $\gamma$-ray variability on intra-day timescale is frequently observed in blazars \citep[e.g.,][]{aharonian07, bonnoli11, aleksic14}, the accompanying high amplitude (3 mags) optical variability is rare \citep[e.g., CTA~102,][]{dammando19}. 

With these overlapping optical and $\gamma$-ray observations and their rich, correlated temporal structures, we can look for temporal offsets between the variations at different energies. We did this using both Javelin \citep{zu2011} and the Interpolated Cross Correlation Function (ICCF) \citep{peterson98, peterson04} methods, focusing on the Javelin results since they are generally less biased and provide better uncertainty estimates \citep{yu2020}. For the inter-optical bands, we used the multi-band LCOGT light curves measured between MJD 58124--58360 (Figure~\ref{fig:tess}, top panel), to find lags between the LCOGT BV, Vr, and ri light curves of $-0.07_{-0.69}^{+0.24}$, $0.05_{-0.27}^{+0.15}$, and $0.06_{-0.43}^{+0.10}$~days, respectively. For the $\gamma$-ray-optical band correlation, we first performed the analysis between the long-term Fermi and ASAS-SN g-band light curves between MJD 58100--59000 and around the overall peak of the light curves (MJD 58800--59000) between Fermi and ASAS-SN-g band, then in short periods with either significant flares or higher quality data during the period with LCOGT coverage (MJD 58124--58360) and the TESS segment (MJD 58927--58954). 
The four $\gamma$-optical lags between the long-term Fermi-ASAS-SN g band, Fermi-ASAS-SN g around the overall peak, Fermi-LCOGT B, and Fermi-TESS are measured as $3.3_{-6.3}^{+7.7}$, $0.9_{-3.1}^{+0.1}$, $0.4_{-3.6}^{+3.2}$, and $-2.1_{-1.5}^{+3.1}$ respectively.
Using the ICCF method, we found that the optical light curves are well correlated with no significant inter-band lags, (e.g., $0.00\pm0.35$ days between $B$ and $i$ bands).
For the $\gamma$-ray and the ASAS-SN light curves, we found a lag $0.00\pm0.45$ day using the ICCF method. 
We also confirmed that all of these light curves are significantly correlated (Figure~\ref{fig:iccf}). The fractional amplitudes of many of the optical and $\gamma$-ray flares are quite similar (Figure~\ref{fig:lc}).

\section{Spectral Evolution\label{sec:spec}}
\begin{figure}
    \centering
    \centerline{\includegraphics[width=1.0\textwidth]{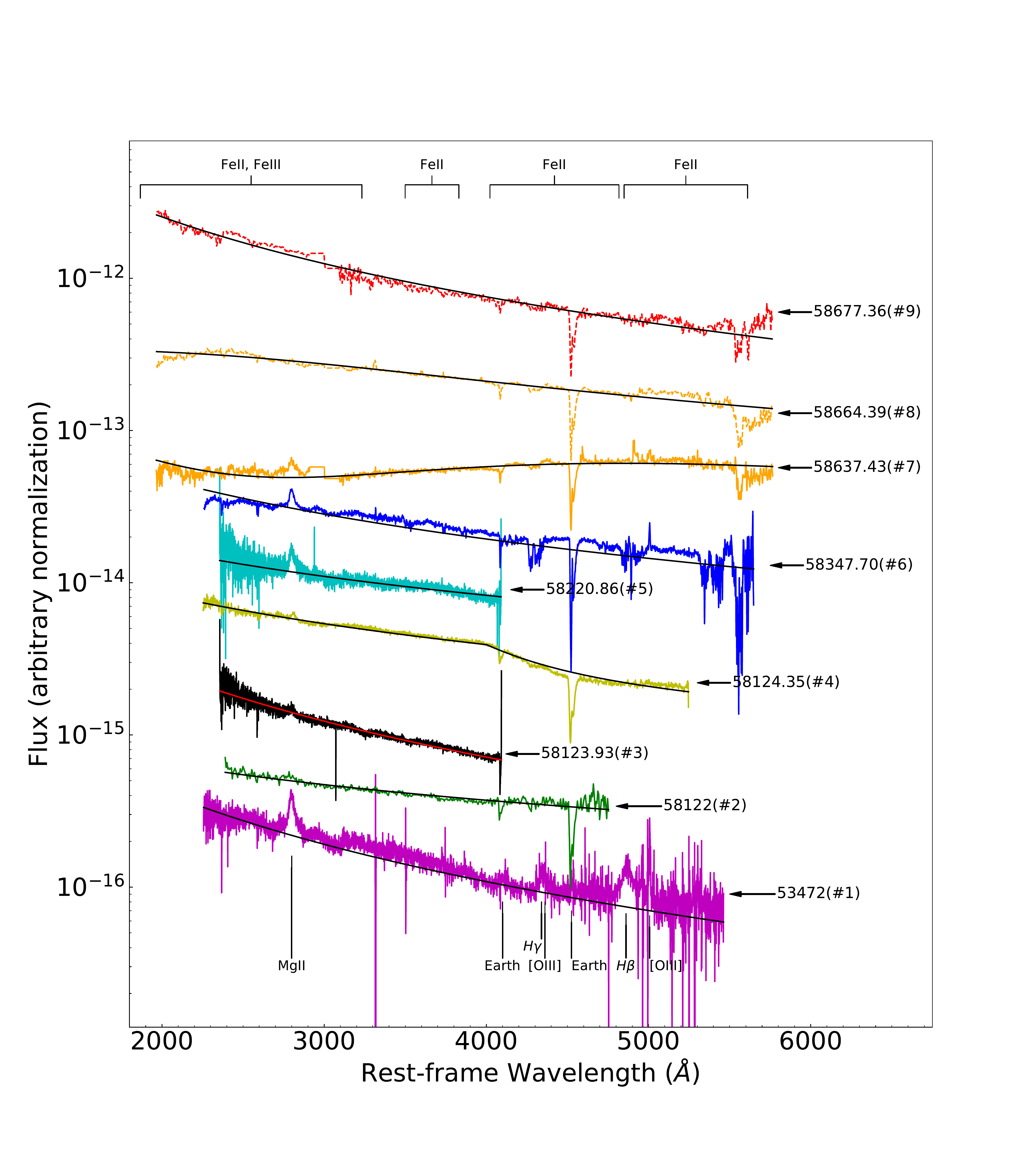}}
    \caption{Spectroscopic evolution of \btwo\ in ascending chronological order from the bottom. The MJD of the observations and the spectrum number are given next to the arrows and the continuum fits are overplotted.}
    \label{fig:spec1}
\end{figure}

\begin{figure}
    \centering
    \centerline{\includegraphics[width=1.0\textwidth]{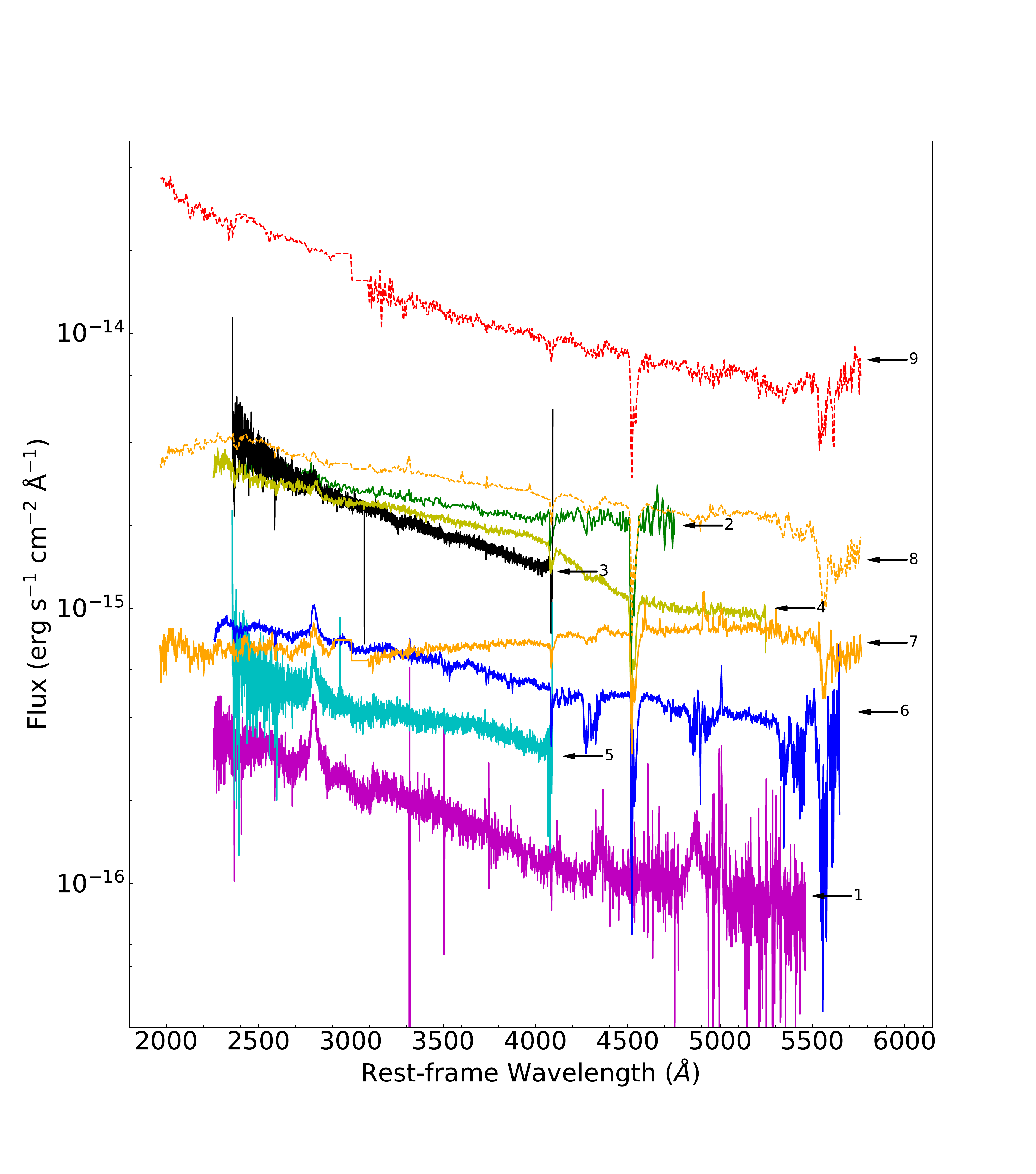}}
    \caption{The spectra in absolute flux units. The number assigned to each spectrum describes the ascending chronological order in Table \ref{tab:table1} and the labels in Fig. \ref{fig:spec1}.}
    \label{fig:spec2}
\end{figure}

We have nine spectra to examine the spectra variability, the archival SDSS spectrum from August 2005 and 8 spectroscopic follow-up observations after the 2018 January outburst. 
The spectra are shown in Figures~\ref{fig:spec1} and \ref{fig:spec2}, where we present the spectra ordered by time in Figure \ref{fig:spec1} and by absolute flux in Figure \ref{fig:spec2}. The first format makes it easier to follow the evolution, while the second makes it easier to see how the spectral structure changes with luminosity. Table~\ref{tab:table1} lists the spectroscopic observations with the parameters describing the continua, and Table~\ref{tab:table2} lists the emission line measurements.
We corrected the spectra for Galactic extinction of $E(B-V) = 0.001$ \citep{schlegel98} and converted them into the rest-frame. 
The spectral analysis was performed using CIAO's Sherpa software \citep{freeman2001}, by minimizing the $\chi^2$ statistics of the fits, which also provides uncertainties of the fitting parameters.
We first fit the continuum by filtering out the spectral regions with major emission lines including \MgII, H$\beta$, H$\gamma$, \OIII\ lines, earth absorption lines, and potential artifacts from data. 
For spectra 1, 6, 7, 8, and 9 with significant Fe emission, we further filtered out Fe emission bands from the continuum fitting process leaving only spectral windows with minimal Fe emission contributions.
However, for spectra 8 and 9, we kept the spectral regions with moderate Fe emission contributions to increase the continuum fitting regions and better constrain the continuum model, since the fitting process suggested more complex continuum models.
For all the spectra, the continuum was fit first using a power law model, since the non-thermal jet emission is assumed to be a power-law, and we obtained reduced $\chi^2$/dof of 1.8/89, 1.5/418, 2.0/4000, 1.1/1722, 0.82/4049, 2.1/3190, 1.64/1791, 1.7/546, and 1.4/348, respectively for the nine spectra.
The reduced $\chi^2$ values are less than 2.1 for all the fits, and we generally consider them to be acceptable, because either the uncertainties of the spectra can be underestimated/overestimated or there are the still unaccounted emission line contributions in the spectral fitting regions.
We next checked if alternative or additional model components are needed for the continuum model, by examining the presence of continuous residuals above or below the best-fit models, and identified spectra 4, 7 and 8.
For spectrum 4, the broken powerlaw model was used to improve the $\chi^2$ value and hence the fit, where the reduced $\chi^2$ is 0.44 significantly decreased from 1.1 for a single power-law. 
For spectra 7 and 8, adding a blackbody component have improved their fitting statistics from reduced $\chi^2 = 1.64$ to 0.94 for spectrum 7 and from 1.7 to 0.63 for spectrum 8. 
The residuals of the fittings were also more randomly scattered about the alternative broken powerlaw model or the addition of blackbody component for spectra 4, 7 and 8, indicating better fits compared to the single powerlaw model.
For the emission line measurements, we followed the general steps of \citet{shen11}, by first fitting a local power-law to the spectral regions containing the emission lines, then adding Gaussian components for the lines. 
We first used one Gaussian model for each emission lines, and obtain reduced $\chi^2 < 1.2$ for all the fits with the residuals randomly scattered about the best-fits, suggesting single Gaussian models are adequate for modeling these lines.
In most cases, we allow the line center, width, and flux to be free parameters. However, when the S/N is too low, we fix the widths of the lines to reasonable prior values. 
For the Fe pseudo-continuum, since the observed profiles can be quite different from those typically observed in non-jetted AGN (Figures~\ref{fig:spec1} and \ref{fig:spec2}), we did not use the template fitting method, but simply estimated the flux by subtracting the continuum from the observed spectra and excluding other known spectral lines.    

\begin{figure}
    \centering
    \centerline{\includegraphics[width=1.2\textwidth]{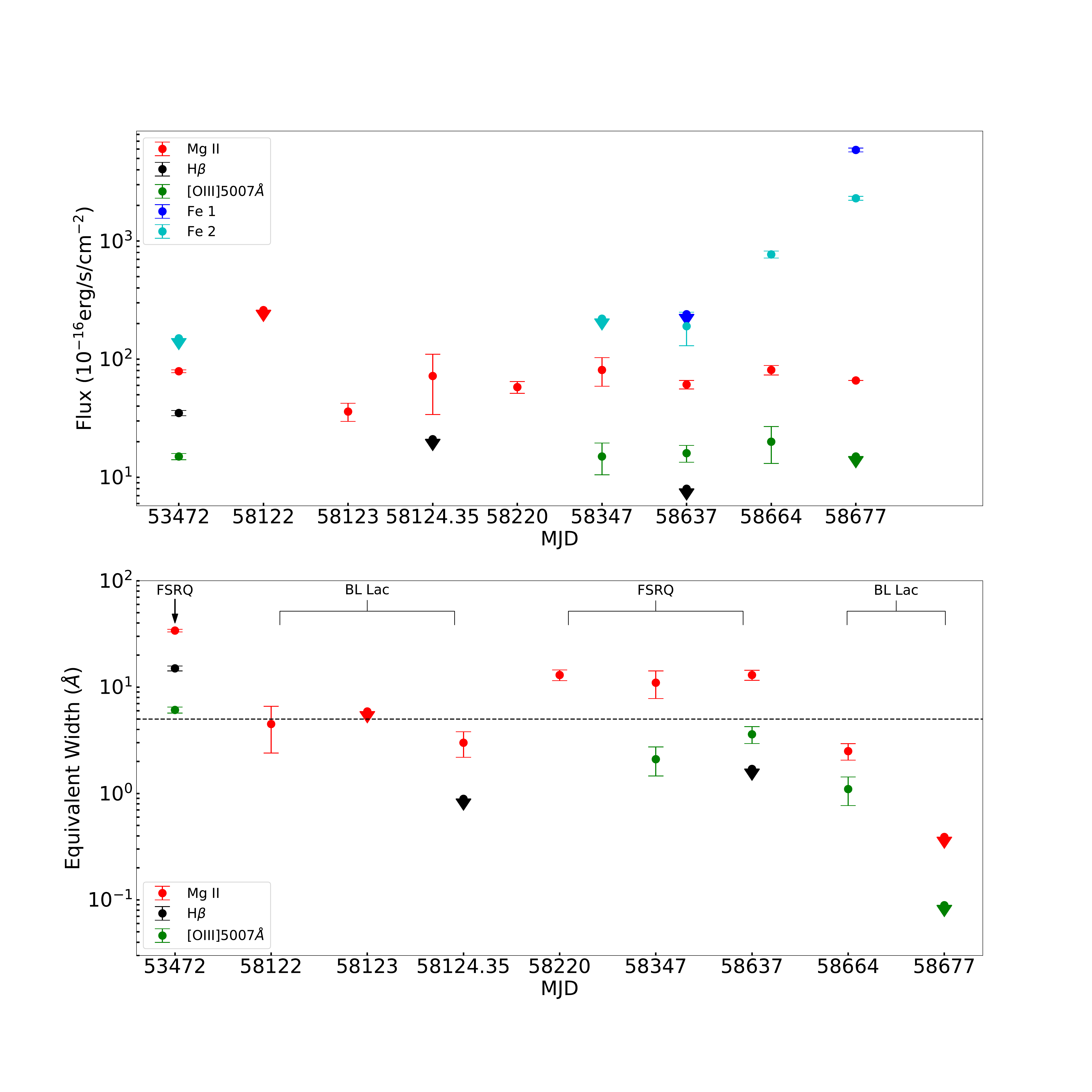}}
    \caption{The evolution of \MgII, H$\beta$, [OIII]5007\AA, and Fe continuum fluxes from spectrum 1 to 9 in the upper panel. Fe 1 corresponds to the Fe II and Fe III emissions in 1250-3100\AA\ and Fe 2 corresponds to 4900-5550\AA\ Fe II emission.  The lower panel shows the evolution of the equivalent widths (see Table \ref{tab:table2}). The black dashed line at 5\AA\ denotes the equivalent width classification between FSRQ and BL Lac states. The downward triangles in both panels are upper limits.}
    \label{fig:spec3}
\end{figure}

\btwo\ shows rich and complex spectral changes, with multiple transitions between the FSRQ and BL Lac spectral types. Here we describe the main features.  From spectrum 1 to 10, we see the source transition from an FSRQ (spec 1) --$>$ BL Lac (specs 2--4) --$>$ FSRQ (spec 5--7) --$>$ BL Lac (specs 8--9), accompanied by complex line and continuum flux changes. Figure~\ref{fig:spec3} shows the line flux and equivalent width variations for the 9 spectral epochs.
The archival SDSS spectrum (spec 1) of \btwo\ is a typical FSRQ spectrum, with a powerlaw continuum and broad \MgII, $H\beta$, $H\gamma$,  and \OIII\ lines with equivalent widths ranging from 4--35\AA.  
Spectra 2--4 were taken during the January 2018 outburst, and we can see that the spectrum evolved into a BL Lac spectrum with an almost featureless continuum and BEL equivalent widths $< 5$\AA\ over spectra 2--4, and then back to an FSRQ in spectrum 5 when the continuum drops.  
The broad lines vary in both the line flux and equivalent widths.  Comparing the \MgII\ flux and equivalent width variations during the transition of FSRQ --$>$ BL Lac --$>$ FSRQ in spectra 1--5 (excluding spectrum 2 because of its large measurement uncertainties), we find that the \MgII\ flux changes by a factor of two and the equivalent width first decreases by a factor of 10 and then increases by a factor of 4. 
The much larger equivalent width variations suggest that the change to having the spectrum of a blazar is mainly due to the large changes in the continuum flux.
We also correlated \MgII\ equivalent width with the continuum and found a negative correlation using Pearson correlation coefficient, with a correlation coefficient of $-0.5$. We see a decreased \MgII\ equivalent width as the jet contribution increases consistent with model predictions \citep[e.g.][]{foschini2012}.
Spectrum 4 exhibits a broken power-law continuum. 

We continued spectroscopic monitoring as the source continued to show large gamma-ray and optical variability.  Spectrum 6 shows a significant \FeII\ pseudo-continuum, and spectra 7 and 8 show additional components that can be modeled with blackbodies with temperatures of 5200 and 12,000~K, respectively, on top of the power-law continuum.
In spectrum 9, the continuum returned to a single power-law, with the addition of Fe pseudo-continuum emission in the rest-frame ultraviolet.
Between spectra 6 and 9, the source again changes from an FSRQ into a BL Lac. 
The broad \MgII\ line flux drops by a factor of 1.7 while the equivalent width drops by a factor of 28.
Figure \ref{fig:spec3} shows the evolution of emission line fluxes and equivalent widths for the major emission lines and Fe emission.
The broad H$\beta$ and H$\gamma$ lines are only marginally detected in spectra 2--8. The upper limits are best constrained in spectrum 7, where H$\gamma$ and H$\beta$ fluxes drop by a factor of 4. The equivalent widths drop by a factor of 9 from the SDSS spectrum.
There are detections of a narrow \OIII5007\AA\ line in spectra 1--9. The line flux increases by a factor of 1.5 between the minimum and maximum values, while the equivalent width changes by a factor of 70. We note that while this small line flux change could be explained by observing conditions like clouds \citep{Fausnaugh2017}, the large change in equivalent width suggests this may be a real phenomenon and not just systematics.
The variations in the \MgII, H$\beta$, H$\gamma$, and \OIII\ lines are consistent with the picture that the differences between FSRQ and BL Lac spectra are due to the changes in the continuum flux.
Where they can be measured, the actual broad line widths and fluxes change little. For example, the FWHM of \MgII\ is $\sim4000$~\kms\ and that of \OIII5007\AA\ is $\sim500$~\kms\ 
both before and after the FSRQ --$>$ BL Lac --$>$ FSRQ evolution.

\begin{deluxetable}{ccclccccr}[b!]
\renewcommand{\arraystretch}{1.5}
\tabletypesize{\footnotesize}
\tablecaption{B2 1420+32 spectra continuum analysis \label{tab:table1}}
\tablecolumns{9}
\tablenum{1}
\tablewidth{0pt}
\tablehead{
\colhead{Spec} & 
\colhead{Telescope} &
\colhead{Resolution} &
\colhead{MJD} &
\colhead{Powerlaw} &
\colhead{Powerlaw} & 
\colhead{Powerlaw} &
\colhead{Blackbody T} &
\colhead{Blackbody} \\
\colhead{} &
\colhead{} &
\colhead{(\AA)} &
\colhead{} &
\colhead{Amplitude\tablenotemark{a}} &
\colhead{Index 1} &
\colhead{Index 2} &
\colhead{(Kelvin)} &
\colhead{Flux\tablenotemark{b}} 
}
\startdata
1 & SDSS & 2.5 & 53472 & $2.3_{-0.0042}^{+0.0042}$ & $1.869_{-0.008}^{+0.008}$ & \nodata & \nodata & \nodata \\
2 & LTLT & 3 & 58122 & $28_{-0.086}^{+0.086}$  & $0.82_{-0.01}^{+0.01}$  & \nodata & \nodata & \nodata \\
3 & 2.4-m MDM & 3 & 58123.93 & $25_{-0.021}^{+0.021}$ & $1.89_{-0.01}^{+0.01}$ & \nodata & \nodata & \nodata \\
4 & 2.16-m Xinglong & 3 & 58124.35 & $23.6_{-0.11}^{+0.11}$ & $1.19_{-0.02}^{+0.02}$ & $2.78_{-0.04}^{+0.04}$ & \nodata & \nodata \\
5 & 2.4-m MDM & 3 & 58220.86 & $4.4_{-0.011}^{+0.011}$ & $0.99_{-0.01}^{+0.01}$  & \nodata & \nodata & \nodata \\
6 & f-JD-Palomar & 7 & 58347.70 & $7.1_{-0.033}^{+0.033}$ & $1.64_{-0.01}^{+0.01}$  & \nodata & \nodata & \nodata \\
7 & SNIFS & 7 & 58637.43 & $4.7_{-0.097}^{+0.095}$ & $1.17_{-0.06}^{+0.06}$ & \nodata & $5200_{-29}^{+29}$ & $4.8_{-0.12}^{+0.12}$ \\
8 & SNIFS & 7 & 58664.39 & $19_{-1.4}^{+1.3}$ & $0.45_{-0.11}^{+0.10}$ & \nodata & $12000_{-380}^{+320}$ & $6.2_{-0.35}^{+0.35}$ \\
9 & SNIFS & 7 & 58677.36 & $160_{-0.34}^{+0.34}$ & $1.8_{-0.0085}^{+0.0085}$  & \nodata & \nodata & \nodata \\
\enddata
\tablenotetext{a}{Normalized to 3000\AA\ with a unit of $10^{-16}$\lumin \cmsq \AA$^{-1}$.}
\tablenotetext{b}{The flux unit is $10^{-12}$\lumin \cmsq.}
\end{deluxetable}

\begin{splitdeluxetable*}{ccccccccccBcccccccccc}
\renewcommand{\arraystretch}{1.5}
\tabletypesize{\footnotesize}
\tablecaption{B2 1420+32 spectra emission line analysis \label{tab:table2}}
\centering
\tablecolumns{20}
\tablenum{2}
\tablewidth{2pt}
\colnumbers
\tablehead{
\colhead{Spec} & 
\colhead{\MgII} &
\colhead{FWHM} &
\colhead{EQW} &
\colhead{H$\beta$} &
\colhead{FWHM} & 
\colhead{EQW} &
\colhead{H$\gamma$} &
\colhead{FWHM} &
\colhead{EQW} &
\colhead{[OIII]4363\AA} &
\colhead{FWHM} &
\colhead{EQW} &
\colhead{[OIII]5007\AA} &
\colhead{FWHM} &
\colhead{EQW} &
\colhead{FeII,FeIII} &
\colhead{FeII} &
\colhead{FeII} &
\colhead{FeII}\\
\colhead{} &
\colhead{Flux} &
\colhead{(km/s)} &
\colhead{(\AA)} &
\colhead{Flux} &
\colhead{(km/s)} &
\colhead{(\AA)} &
\colhead{Flux} &
\colhead{(km/s)} &
\colhead{(\AA)} &
\colhead{Flux} &
\colhead{(km/s)} &
\colhead{(\AA)} &
\colhead{Flux} &
\colhead{(km/s)} &
\colhead{(\AA)} &
\colhead{1250-3100\AA} &
\colhead{3530-3800\AA} &
\colhead{4070-4750\AA} &
\colhead{4900-5550\AA}
}
\startdata
1 & $79_{-2.1}^{+2.1}$ & $4800_{-160}^{+160}$ & $34_{-0.97}^{+0.97}$ &
$35_{-1.7}^{+1.8}$ & $3800_{-210}^{+220}$ & $15_{-0.77}^{+0.81}$ & $9.9_{-1.6}^{+1.7}$ & $2600_{-310}^{+340}$ & $4.3_{-0.70}^{+0.75}$ & $2.2_{-0.39}^{+0.39}$ & $480^{*}$ & $0.96_{-0.17}^{+0.17}$ & $14_{-0.87}^{+0.90}$ & $470_{-38}^{+41}$ & $6.1_{-0.39}^{+0.40}$ & \nodata & $30_{-42}^{+42}$ & \nodata & $<150$\\
2 & $<260$  & $<1400$ & $4.5_{-2.1}^{+2.1}$ &
\nodata & \nodata & \nodata & \nodata & \nodata & \nodata & \nodata & \nodata & \nodata & \nodata & \nodata & \nodata & \nodata & \nodata & \nodata & \nodata \\
3 & $36_{-6.3}^{+6.3}$ & $3000^{*}$ & $<5.9$ & \nodata & \nodata & \nodata & \nodata & \nodata & \nodata & \nodata & \nodata & \nodata & \nodata & \nodata & \nodata & \nodata & \nodata & \nodata & \nodata \\ 
4 & $72_{-35}^{+38}$ & $3200_{-1300}^{+2000}$ & $3.0_{-0.72}^{+0.81}$ & $<21$ & $2000^{*}$ & $<0.89$ & $<11$ & $2000^{*}$ & $<0.47$ & $<3.3$ & $< 210$ & $<0.15$ & \nodata & \nodata & \nodata & \nodata & \nodata & \nodata & \nodata \\
5 & $58_{-6.2}^{+6.6}$ & $3800_{-430}^{+490}$ & $13_{-1.4}^{+1.5}$ & \nodata & \nodata & \nodata & \nodata & \nodata & \nodata & \nodata & \nodata & \nodata & \nodata & \nodata & \nodata & \nodata & \nodata & \nodata & \nodata \\
6 & $81_{-19}^{+22}$ & $3600_{-890}^{+1200}$ & $11_{-2.7}^{+3.2}$ & \nodata & \nodata & \nodata & \nodata & \nodata & \nodata & $<2.5$ & $< 480$ & $<0.35$ & $15_{-4.5}^{+4.2}$ & $< 420$ & $2.1_{-0.64}^{+0.60}$ & \nodata & $<$ 160 &  $<$ 110 & $< 220$ \\
7 & $61_{-4.9}^{+5.1}$ & $4800_{-530}^{+580}$ & $13_{-1.3}^{+1.4}$ & $<8.0$ & $2000^{*}$ & $<1.7$ & $2.3_{-1.7}^{+1.7}$ & $2000^{*}$ & $0.49_{-0.37}^{+0.37}$ & $<0.67$ & $< 480$ & $<0.16$ & $16_{-2.4}^{+2.6}$ & $610_{-103}^{+130}$ & $3.6_{-0.60}^{+0.65}$ & $<240$ & \nodata & \nodata & $190_{-60}^{+60}$ \\
8 & $48_{-13}^{+14}$ & $2000_{-550}^{+680}$ & $2.5_{-0.45}^{+0.44}$ & \nodata & \nodata & \nodata & \nodata & \nodata & \nodata & $<2.4$ & $< 480$ & $<0.13$ & $20_{-6.9}^{+6.9}$ & $< 420$ & $1.1_{-0.33}^{+0.33}$ & \nodata & \nodata & \nodata & $770_{-52}^{+52}$ \\
9 & $<66$ & $1000^{*}$ & $<0.39$ & \nodata & \nodata & \nodata & \nodata & \nodata & \nodata & $<13$ & $< 480$ & $<0.09$ & $<15$ & $< 420$ & $<0.089$ & $5900_{-220}^{+220}$ & \nodata & \nodata & $2300_{-87}^{+87}$ \\ 
\enddata
\tablenotetext{a}{Flux unit is $10^{-16}$\lumin \cmsq.}
\tablenotetext{*}{These parameters were fixed.}
\end{splitdeluxetable*}

\section{Discussion\label{sec:dis}}
Our multi-wavelength and spectroscopic monitoring observations show that \btwo\ exhibits extreme spectral and temporal variability. We observe flux increases over the past two decades by factors of 1500 (8 mags) and 100 (5 mags) in the $\gamma$-ray and optical bands, respectively, with correlated optical and $\gamma$-ray variability.  
The $\gamma$-ray and optical flux changes can be up to factors of 40 and 16 respectively, on week-to-month timescales and a factor of 3 on intraday timescales in the optical. The optical and $\gamma$-ray lightcurves are well-correlated with lags $<$ 3 days.

We can estimate the sizes of the $\gamma$-ray and optical emission regions based on our variability and lag measurements. Here we use the mass reported in \citet{brotherton15} ($M_{BH} \simeq 4 \times 10^8 M_\odot$), implying an Eddington luminosity of $5.2\times10^{46}~\lumin$.
We also measured the black hole mass independently using H$\beta$ line width and luminosity from the SDSS spectrum and found the mass to be consistent within 2\% the above mentioned value, and the \MgII\ mass is within 40\%.  
The black hole has a gravitational radius size of $r_g = 5.9\times10^{13}$ cm.
Assuming a typical Doppler factor of $\delta=10$ for the jet \citep{hovatta09, liodakis17} and considering the source redshift of  $z=0.68$, $\Delta t_{intr} = \Delta t_{obs} \delta / (1+z)$,  an observed lag of one day corresponds to an emission region size of 260~$r_g$.  The measured inter-optical lags are $<$ 0.5 days, corresponding to an intrinsic source size of $<$ 130~$r_g$.  Using a conservative lag uncertainty of 3 days for our $\gamma$-ray-optical lag measurements on short time-scales, the $\gamma$-ray and optical emission regions are separated by $<$ 800~$r_g$.

Dramatic spectral variations were also observed. In particular, we observe, for the first time, multiple, rapid transitions between the FSRQ and BL Lac spectral classifications. Few changing-look blazars have been reported previously, for example VRO 42.22.01 \citep{vermeulen95} and 5BZB~J0724+2621 \citep{ac16}, where a transformation from a BL Lac to an FSRQ spectral type was observed once. 
For our source, the initial FSRQ spectrum with broad emission lines with \MgII, H$\beta$, and H$\gamma$ evolves to the featureless spectrum of a BL Lac object, and then back again, with the reappearance of \MgII\ lines plus a new \FeII\ and \FeIII\ pseudo-continuum and other continuum features. However, the Balmer emission lines are never significantly detected after the first flares, except for H$\gamma$ in the last spectrum.
The optical continuum changes in shape, where we can model it as a single powerlaw, a broken powerlaw, or a powerlaw plus blackbody components, depending on the spectrum.

Our optical spectra show that the optical emission during flares is still dominated by a powerlaw continuum, presumably from the jet. A jet origin is particularly indicated by the broken powerlaw spectrum, which is a characteristic non-thermal emission feature and has never been observed from accretion disks \citep[][]{gierlinski2001,wu2013}.

The large change in some of the line features shows that the BLR is significantly affected by the $\gamma$-ray and optical flares. While the \MgII, \OIII, and Balmer line fluxes vary by a factor of 2--3, the equivalent width changes can be as high as a factor of 150 because of the huge changes in the optical continuum flux.
This is consistent with the less dramatic case of 3C~279, where the \lya\ flux is observed to vary by a factor of $\sim$2, while the continuum changed by a factor of up to 50 \citep{koratkar98}.
The lower variability amplitudes observed in these lines corroborate with the conclusions from studies of larger samples of moderate continuum variability blazars that the BLR clouds are mainly photo-ionized by the accretion disk with significant contribution from the jet to the ionization \citep[e.g.,][]{isler13, isler15}. 
The relative consistency in the \MgII\ and \OIII\ line width measurements also suggests that the BLR is only partially affected by the dramatic optical and $\gamma$-ray variability.
The appearance of a \FeII\ and \FeIII\ pseudo--continuum is the exception, where we observe a flux increase by a factor of 45 from the archival SDSS to the most recent spectrum (spec 9), with a peak flux of 3\% Eddington luminosity.
The non-detections of Fe pseudo--continuum in spectra 2--5 can be caused by the reduction of the equivalent widths by the increase of the continuum flux.
The appearance of a strong \FeII\ and \FeIII\ pseudo-continuum suggests the disruption of dust clouds by shocks or radiation, which would free up a large amount of Fe \citep[e.g.,][]{kishimoto2011, baskin2018}. 
The variability in the emission line fluxes of different species (typically a factor of 2) and \FeII\ and \FeIII\ ($\sim$45) suggests energy transfers from the relativistic jet to sub-relativistic components. 
It is also possible that the variations in the continuum flux from the disk \citep[e.g.,][]{kelly09,macleod10} could drive dust destruction while being masked by the far larger variations in the jet component. 

One optical spectrum (spec 7) shows a prominent continuum feature, which is well-fit by a 5,200~K blackbody, and a second spectrum (spec 8) shows a prominent component well-fit by an 12,000~K blackbody. The two components have luminosities of 18--24\% $L_{Edd}$.
The narrowness of the blackbody peaks suggests that the emission source is most likely sub(mildly)-relativistic, because relativistic Doppler effects will broaden any narrow features.
A modification from the powerlaw jet emission combined with the beaming effect can mimic a single blackbody spectral shape.
These single temperature blackbody components are difficult to interpret as radiation from the accretion disk, because accretion disks span broad temperature ranges, leading to UV/optical SEDs that are essentially power-laws.
Blackbody-like spectral components have been observed in blazars and they are commonly interpreted as the host galaxy contribution, particularly since many of them show absorption features typical of host galaxies \citep[e.g.,][]{paiano20}.
However, the blackbody components detected in \btwo\ are clearly not from the host because the host contribution is constant. Here, the blackbody components are detected only when the source is near peak brightness, while there is no significant host component visible even in the archival, low-state, SDSS spectrum.

The unique blackbody components could be from the jet itself, if the jet is precessing and we are occasionally observing a part of the jet with low Doppler beaming factors. This model of changing viewing angles was also proposed to interpret the huge $\gamma$-ray and optical flux changes in CTA~102 \citep{dammando19}.
Alternatively, it is possible that there are changes in the opening angles of the jet, and the blackbody component can be from episodes of jet activity with larger opening angles and low Lorentz factors propagating through a dust-rich region (presumably the torus) to free-up the ions producing the Fe pseudo-continuum.
Regardless of the interpretation, the low Doppler factor suggests that these narrow blackbody spectra are more representative of the jet spectrum seen at a typical location in the central engine, and not directly along the jet.

AGN feedback has been broadly classified into the ``quasar mode'' and ``radio mode''. 
The ``quasar mode'' is feedback from either the radiation (high-Eddington regime) or disk winds in the non-Eddington regime, while the ``radio mode'' is kinetic feedback from decelerated radio jet/lobes in low luminosity radio galaxies or galaxy clusters.  
Strongly relativistic jets are seldom considered as important galaxy scale feedback sources, because they penetrate through the galaxy and are only decelerated to mildly relativistic speeds for larger (cluster) scales.
Here, we show that these jets may drive intermittent sub(mildly)-relativistic shocks in the central engine/host galaxy with luminosities of 20\% $L_{Edd}$ or Fe emission flux changes of 5\% $L_{Edd}$.

Finally, we summarize the main conclusions of this paper:  
\vspace{-1.5mm}
\begin{itemize}
    \item Between 2016--2019, the $\gamma$-ray and optical fluxes increased by factors of 1500 (8 mags) and 40 (4 mags) respectively. The optical variability amplitude observed is unprecedented, with the optical flux increasing by a factor of 100 (5 mags) compared to the SDSS observations in 1995.
    
    \item The optical-$\gamma$-ray and inter optical band correlations constrain the $\gamma$-ray-optical lag to be $<$ 3 days and inter-optical band lags to be $<$ 0.5 days, corresponding to emission distance/sizes of less than $\sim800$$r_g$ and $\sim130$$r_g$. 
    
    \item \btwo\ is a changing-look blazar, transiting between the two major classifications of blazars, the FSRQ and BL Lac categories due to dramatic changes in the jet continuum flux diluting the line features.
    
    \item Complex spectral evolution is observed in both the continuum and emission lines, suggesting dramatic changes in the jet and photoionization properties of the emission line regions.  The emergence of strong \FeII\ and \FeIII\ pseudo-continuum is consistent with the sublimation of dust grains by either radiation or shocks releasing more Fe ions into the broad line regions. The Fe line fluxes approach 3\% $L_{Edd}$.
    
    \item For the first time, we detect components in the optical spectra
    consistent with single temperature blackbody emission, with 20\% of the Eddington luminosity.  
\end{itemize}
This extreme variability we describe here has not been observed before. However, it may not be uncommon, because dedicated multi-band and spectroscopic monitoring of blazars are still rare. Dedicated searches for more changing-look blazars will extend the changing-look AGN studies to jetted AGNs and allow us to utilize the dramatic spectral changes to reveal AGN/jet physics.

\acknowledgments
We thank B.\ Peterson, R.\ Pogge, and J. Zhang for helpful discussions.  HDM\ and XD\ acknowledge the financial support from the NASA ADAP program NNX17AF26G.
We thank the Las Cumbres Observatory and its staff for its continuing support of the ASAS-SN project. ASAS-SN is supported by the Gordon and Betty Moore Foundation through grant GBMF5490 to the Ohio State University, and NSF grants AST-1515927 and AST-1908570. Development of ASAS-SN has been supported by NSF grant AST-0908816, the Mt. Cuba Astronomical Foundation, the Center for Cosmology and AstroParticle Physics at the Ohio State University, the Chinese Academy of Sciences South America Center for Astronomy (CAS-SACA), and the Villum Foundation. BJS, CSK, and KZS are supported by NSF grant AST-1907570. BJS is also supported by NASA grant 80NSSC19K1717 and NSF grants AST-1920392 and AST-1911074. CSK and KZS are supported by NSF grant AST-181440. KAA is supported by the Danish National Research Foundation (DNRF132). MAT acknowledges support from the DOE CSGF through grant DE-SC0019323. Support for JLP is provided in part by FONDECYT through the grant 1151445 and by the Ministry of Economy, Development, and Tourism's Millennium Science Initiative through grant IC120009, awarded to The Millennium Institute of Astrophysics, MAS. TAT is supported in part by Scialog Scholar grant 24215 from the Research Corporation.
We acknowledge the Telescope Access Program (TAP) funded by NAOC, CAS, and the Special Fund for Astronomy from the Ministry of Finance. PJV is supported by the National Science Foundation Graduate Research Fellowship Program Under Grant No. DGE-1343012. 

\vspace{5mm}
\facilities{ASAS-SN, Fermi, TESS, LCOGT, LTLT, MDM Hiltner, BAO 2.16m, Palomar, Hawaii 88in, Swift, SDSS}

\begin{deluxetable}{cclccccr}[b!]
\tablecaption{B2 1420+32 optical lightcurves \label{tab:lctable1}}
\tablecolumns{8}
\tablenum{3}
\tablewidth{0pt}
\tablehead{
\colhead{MJD} & 
\colhead{Telescope} &
\colhead{Band} &
\colhead{Magnitude} &
\colhead{Uncertainty} 
}
\startdata
53479.27321 & CRTS & V & 18.25 & 0.12 \\
53479.28138 & CRTS & V & 18.48 & 0.13 \\
53479.28951 & CRTS & V & 18.31 & 0.12 \\
53479.2977 & CRTS & V & 18.41 & 0.13 \\
53562.16491 & CRTS & V & 18.25 & 0.12 \\
53562.17136 & CRTS & V & 18.14 & 0.12 \\
58100.00768 & ASAS-SN & g & 17.81 & 0.20 \\
58104.00069 & ASAS-SN & g & 17.75 & 0.14 \\
58115.98322 & ASAS-SN & g & 15.52 & 0.03 \\
58116.97997 & ASAS-SN & g & 16.17 & 0.03 \\
58118.02747 & ASAS-SN & g & 15.62 & 0.03 \\
58119.02784 & ASAS-SN & g & 15.56 & 0.03 \\
58124.0043 & LCOGT & B & 16.41 & 0.06 \\
58125.0281 & LCOGT & B & 16.36 & 0.07 \\
58125.9248 & LCOGT & B & 16.70 & 0.07 \\
58128.0192 & LCOGT & B & 16.88 & 0.07 \\
58128.9496 & LCOGT & B & 16.73 & 0.12 \\
58141.917 & LCOGT & B & 17.76 & 0.07 \\
58927.60 & TESS &  & 14.36 & 0.00 \\
58927.62 & TESS &  & 14.35 & 0.00 \\
58927.64 & TESS &  & 14.29 & 0.00 \\
58927.67 & TESS &  & 14.31 & 0.00 \\
58927.69 & TESS &  & 14.30 & 0.00 \\
58927.71 & TESS &  & 14.25 & 0.00 \\
\enddata
\tablenotetext{}{This is presented for form and content. The full table is available in the ApJ online version of the paper.}
\end{deluxetable}

\begin{deluxetable}{cclccccr}[b!]
\tablecaption{B2 1420+32 Fermi LAT lightcurves \label{tab:lctable2}}
\tablecolumns{8}
\tablenum{4}
\tablewidth{0pt}
\tablehead{
\colhead{MJD} & 
\colhead{Bin size} &
\colhead{Flux} &
\colhead{Uncertainty} 
}
\startdata
54772.66 & 2 months & 0.72 & 0.05 \\
54832.66 & 2 months & 0.29 & 0.30 \\
54892.66 & 2 months & 1.22 & 0.58 \\
55012.66 & 2 months & 0.23 & 0.05 \\
55072.66 & 2 months & 1.24 & 1.06 \\
55132.66 & 2 months & 0.81 & 0.55 \\
57504.0 & 3 days & 5.57 & 0.48 \\
57543.0 & 3 days & 9.86 & 1.82 \\
57567.0 & 3 days & 7.41 & 5.35 \\
57585.0 & 3 days & 7.79 & 3.71 \\
57609.0 & 3 days & 6.06 & 3.70 \\
57615.0 & 3 days & 2.12 & 1.45 \\
58104.5 & 1 day & 12.5 & 11.5 \\
58106.5 & 1 day & 28.1 & 13.0 \\
58110.5 & 1 day & 19.5 & 11.3 \\
58112.5 & 1 day & 11.4 & 4.83 \\
58114.5 & 1 day & 10.6 & 4.98 \\
58115.5 & 1 day & 13.8 & 7.08 \\
58927.86 & 0.5 day & 53.8 & 16.4 \\
58928.36 & 0.5 day & 40.4 & 19.2 \\
58928.86 & 0.5 day & 37.1 & 11.0 \\
58929.36 & 0.5 day & 49.4 & 3.63 \\
58929.86 & 0.5 day & 55.4 & 15.9 \\
58930.36 & 0.5 day & 32.3 & 14.9 \\
\enddata
\tablenotetext{}{This is presented for form and content. The full table is available in the ApJ online version of the paper.}
\tablenotetext{}{The flux is in the units of 10$^{-8}$ \pflux.}
\end{deluxetable}

\end{document}